\documentstyle[preprint,aps]{revtex}
\begin{document}
\preprint{}
\draft
\tighten
\title{The D1-D5 brane system in six dimensions}

\author{Marika Taylor-Robinson \footnote{E-mail:
    mmt14@damtp.cam.ac.uk}}
\address{Department of Applied Mathematics and Theoretical Physics,
\\University of Cambridge, Silver St., Cambridge. CB3 9EW}
\date{\today}
\maketitle

\begin{abstract}
{We consider scattering of minimal coupled scalars from a six-dimensional 
black string carrying one and five brane charges but no Kaluza-Klein 
momentum. The leading correction to the absorption cross section is found 
by improved matching of inner and outer solutions to the wave equation. 
The world sheet interpretation of this correction follows from the breaking 
of conformal invariance by irrelevant Born-Infeld corrections. We note that 
discrepancies in normalisation are caused by there being
two effective length scales in the black string geometry but only
one in the effective string model and comment on 
the implications of our results for the effective string model. 
}

\end{abstract}
\pacs{PACS numbers: 04.50.+h, 04.65.+e}
\narrowtext

\section{Introduction}
\noindent

There has been a great deal of progress over the last year in 
understanding the strong coupling limit of large $N$ gauge theory 
\cite{G_K_P}, \cite{K_1}, \cite{Kl_Gub}, \cite{G_K_T}, \cite{P_1} . 
Of particular importance is 
Maldacena's conjecture \cite{Mald_1} that the world volume 
theories of certain 
coincident branes are related to string theory or M theory on 
backgrounds consisting of anti-de Sitter spaces times spheres. Subsequent 
work in \cite{PoKl} and \cite{W_1} elaborated on the precise nature 
of this correspondence and many more papers on the subject have followed. 
Much of the interest has been focussed on the D3-brane system and 
on the D1-D5 brane system with momentum along the string direction. In this 
paper we will be interested in the latter system, although to simplify the 
calculations from both the world sheet and the supergravity points of
view we shall mostly consider the zero momentum extremal system, which 
corresponds to the zero temperature limit of the effective string model. 

The relationship between correlators in the world sheet theory and
low energy absorption in the entire black string or black hole metric has been 
extensively discussed in the literature. The first work in this direction 
appeared in \cite{CM} and was followed by papers demonstrating the precise
agreement between the semi-classical and world sheet calculations
of the minimal scalar absorption cross section \cite{DM1}, \cite{DM2} 
for extreme black 
holes. In \cite{MS} the calculation was extended to near extreme black holes 
within the so-called dilute gas regime, and many further papers have 
explored scattering in other regions of the moduli space, including 
\cite{KM}, \cite{Mat_1}, \cite{HM} and \cite{Do}. Recent discussion of
the five dimensional black hole system in the context of the the
relationship between the conformal field theory and anti-de Sitter
supergravity can be found in \cite{S_de_A}.  

\bigskip

As Maldacena and others have pointed out, if one 
takes the throat limit of the black string solution 
one can identify the part of the metric which 
determines the correlation functions in the conformal limit. However the 
string geometry is only anti-de Sitter out to a characteristic radius $R$ 
which is related to the number of D1 and D5 branes. When one specifies the 
world sheet theory with the DBI action powers of the string scale $\alpha'$ 
suppress the non renormalisable interactions. Even at strong coupling, when 
the string geometry is smooth, the corrections are detectable in the 
energy dependence of the absorption cross section. Such departures from 
world volume conformal invariance were discussed in detail for the D3-brane in 
\cite{Kle_Has} and our discussion parallels theirs in many
respects. Related discussions on the absorption of two form
perturbations by three-branes are to be found in \cite{Raj} and
absorption by extremal three branes is discussed generally in \cite{Mat_mat}.

From the supergravity point of view we look for corrections to the absorption
cross section of minimal scalars by matching the solutions to the wave 
equation more carefully between inner and outer regions. Just as in 
\cite{Kle_Has} we find that the wave equation has a self-dual point, at 
precisely the radius of the effective anti-de Sitter space. Improved matching
leads us to find a leading order logarithmic correction to the absorption
cross section. One would expect a correction of the same type for scattering 
within the related five dimensional black hole carrying three charges. 

\bigskip

The logarithmic term encodes the leading order departure from the conformal 
limit. This breakdown can be interpreted in terms of the effective
world sheet action for the string: non renormalisable interactions 
enter the action at subleading order. We look at the effect of
subleading couplings of minimally coupled scalars to
operators of conformal dimension four and higher on the string
world sheet. We find that such couplings allow us to reproduce the 
logarithmic form of the leading corrections to the absorption
probability. 

However the normalisations of the corrections predicted by
the effective string model do not agree with those found
semi-classically. The corrections will not agree unless we extend 
the effective world sheet (or worldvolume) theory to take 
account of both the five
brane and one brane charges. Put another way, there are two length 
scales in the black string geometry, related to the two distinct
charges, and any effective worldvolume model would have to take
account of both. 

Similarly, although one can predict the leading order cross section
for a massless scalar using the duality of the near horizon geometry
to a boundary conformal field theory, one cannot successfully predict 
corrections. The near horizon geometry depends only on one parameter,
which appears in the dual conformal field theory, whereas the
supergravity calculation depends on two.  

\bigskip

The plan of the paper is as follows. In \S \ref{ii} we discuss the
supergravity background describing a black string carrying two charges
in six dimensions and in \S \ref{iii} we consider the
supergravity analysis of scattering from the string. In \S \ref{iv} we
consider the effective string analysis whilst in \S \ref{v} we consider the
dual conformal field theory. We present our conclusions in \S \ref{vi}.   

\section{The black string spacetime} \label{ii}
\noindent

The low energy effective action for ten dimensional type IIB string
theory contains the terms
\begin{equation}
\frac{1}{2 \kappa_{10}^2} \int d^{10} x \sqrt{-G} \left [ e^{-2\Phi} \left (R
    + 4 (\partial \Phi)^2 \right ) - \frac{1}{12} H^2 \right],
\end{equation}
where $\Phi$ is the dilaton and $H$ is the RR three form field strength. 
The ten-dimensional solution in which we are interested is 
\begin{equation}
ds^2 = (H_1 H_5)^{-\frac{1}{2}} \left [ -dt^2 + dx^2 + H_1
  \sum_{i=6}^{9} dx_i^2 \right ] + 
(H_1 H_5)^{\frac{1}{2}} [ dr^2 + r^2 d\Omega_3^2 ],
\end{equation}
where we give the metric in the string frame and the harmonic
functions are given by 
\begin{equation}
H_1 = 1 + \frac{r_1^2}{r^2}, \hspace{10mm} H_5 = 1 + \frac{r_5^2}{r^2}.
\end{equation}
The ten dimensional dilaton is 
\begin{equation}
e^{-2 \Phi} = \frac{H_5}{H_1},
\end{equation}
which implies that when one wraps the five brane on a four torus 
the six dimensional dilaton 
$\Phi_{6} = \Phi - \frac{1}{4} \ln G_{int}$, with $G_{int}$ the
determinant of the metric on the torus, is constant. We will not need
the explicit form of the three form in what follows. The effective six
dimensional action in the Einstein frame is 
\begin{equation}
S_{6} = \frac{1}{2 \kappa_{6}^2} \int d^6x  \sqrt{-g} \left [ R - (\partial
  \Phi_6)^2 + ... \right]. \label{6d_act}
\end{equation} 
The solution for the six-dimensional black string 
in the Einstein frame is then 
\begin{equation}
ds^2 = (H_1 H_5)^{-\frac{1}{2}} (-dt^2 + dx^2) + (H_1
H_5)^{\frac{1}{2}} (dr^2 + r^2 d\Omega_3^2),
\end{equation}
If $g$ is the ten-dimensional coupling strength then 
\begin{equation}
\kappa_{10}^2 = 64 \pi^7 g^2 \alpha'^4.
\end{equation}
Dimensionally reducing on a four torus of volume $V$ the associated
six-dimensional variables are 
\begin{equation}
v = \frac{V}{(2 \pi)^4 \alpha'}; \hspace{10mm} g_6 =
\frac{g}{\sqrt{v}}; \hspace{10mm} \kappa_{6}^2 = 4 \pi^3 g_6^2
\alpha'^2. \label{eq_kap}
\end{equation}
The charges of the black string are given by 
\begin{equation}
r_1^2 = \frac{g \alpha' n_1}{v}; \hspace{10mm} r_5^2 = g \alpha' n_5,
\end{equation}
where $n_1$ and $n_5$ are the number of units of D1-brane and D5-brane
charge respectively. 
In the decoupling limit, we can neglect the constant terms in the
harmonic functions and the metric becomes that of $AdS_3 \times S^3$:
\begin{equation}
ds^2 = \frac{R^2}{z^2} \left [ -dt^2 + dx^2 + dz^2 \right ] + R^2 d\Omega_3^2,
\label{str_met}
\end{equation}
with $z = R^2 /r$ and the radius of the effective anti-de Sitter
space being defined by 
\begin{equation}
R^2 = r_1 r_5 = g_{6} \alpha' \sqrt{n_1 n_5} \label{eq_R}.
\end{equation}
For this system, Maldacena's conjecture \cite{Mald_1}
is that the $(1+1)$-dimensional 
conformal field theory describing the Higgs branch of the D1-D5 brane system 
on the torus is dual to type IIB theory on $(AdS_3 \times S^3)_R \times 
T^4_{(r_1^2/r_5^2)}$, where the subscripts indicate the effective
``radii'' of the manifolds. 

\section{Scattering of minimal coupled scalars: semi-classical
  calculation} \label{iii}
\noindent

We firstly consider scattering of a minimally coupled scalar in the
black string metric. Examples of such scalars include the six
dimensional dilaton, which is constant in the background, and
transverse graviton components. The equation of motion for a mode of
frequency $\omega$ of a  
minimally coupled scalar $\phi$ in the black string metric is 
\begin{equation}
[\frac{1}{r^3} \partial_r (r^3 \partial_r) + \omega^2 H_1 H_5] \phi = 0. 
\end{equation}
The parameter $R^2 = r_1 r_5$ which describes the scale of the anti-de
Sitter space plays an important r\^{o}le in determining the form of
the solutions. We divide the spacetime into inner and outer regions 
defined by $r \ll R$ and $r \gg R$; we will consider low energy scattering
and so assume that $\omega R \ll 1$. 

In the region $r > R$ we
look for a solution of the form $\phi(r) = \psi(\omega r)/r$ 
where $\psi$ satisfies 
\begin{equation}
[\rho^2 \psi'' + \rho \psi' - (1 - \omega^2 Q^2) \psi + \rho^2 \psi] =
- \frac{(\omega^4 R^4)}{\rho^2} \psi, \label{can_eq1}
\end{equation}
with $\rho = \omega r$ and $Q^2 = r_1^2 + r_5^2$. For small $\omega
R$ the leading order solution of this equation is, as first discussed in 
\cite{DM1}, 
\begin{equation}
\psi(\rho) = \alpha J_{\nu}(\rho) + \beta J_{-\nu} (\rho),
\end{equation}
where 
\begin{equation}
\nu = (1 - \omega^2 Q^2)^{\frac{1}{2}}.
\end{equation}
We can regard the right hand side of (\ref{can_eq1}) as a small
perturbation in the outer region $r \gg  R$.
 
In the region $r < R$ there is a natural choice of reciprocal variable
$y = \omega R^2 /r$ in terms of which the wave function $\phi(y) = y f(y)$
satisfies 
\begin{equation}
[y^2 f'' + y f' - (1 - \omega^2 Q^2) f + y^2 f] =
- \frac{(\omega^4 R^4)}{y^2} f. \label{can_eq2}
\end{equation}
For $r \ll R$ the term on the right hand side is negligible and the
leading order solution is thence 
\begin{equation}
f(y) = H^{(2)}_{\nu}(y),
\end{equation}
where we have chosen the solution to be pure infalling at the
horizon. Just as for the D3-brane and M-branes, the equation of motion
for the minimal scalar has a self-dual point defined by the radius $R$ of the
effective anti-de Sitter space. 

Matching the amplitude of $\phi$ to leading order at $r = R$, assuming
that $\nu \approx 1$ one finds that 
\begin{equation}
\alpha = \frac{ 4 i }{\pi \omega},
\end{equation}
with $\beta = 0$. 
Note that although such a naive matching scheme seems invalid since
neither solution holds at $r=R$ it does in fact work. We can
find the solution for the scalar field at the self-dual point as an
expansion in $\omega R$, and then match the leading order term to obtain
this result.  

\bigskip

Now the asymptotic form of the infalling wave function is 
\begin{equation}
\phi(y) = y H_{\nu}^{(2)}(y) \approx \sqrt{\frac{2y}{\pi}} \exp \lbrace i(y -
  \frac{1}{2} \nu \pi - \frac{1}{4} \pi) \rbrace,
\end{equation}
which implies that the ingoing flux defined by 
\begin{equation}
F_{r=0} = \frac{1}{2i} \lbrace \phi^{\ast} r^3 (\partial_r \phi) 
- \phi r^3 (\partial_r \phi^{\ast}) \rbrace |_{r=0}
\end{equation}
is given by 
\begin{equation}
F_{r=0} = \frac{2 \omega^2 R^4}{\pi}.
\end{equation}
Since the ingoing part of the wavefunction at infinity is given by
\begin{equation} 
\phi(r) \approx \alpha \sqrt{\frac{1}{2\pi \omega r^3}} 
\exp \lbrace i(\omega r -  \frac{1}{2} \nu \pi - \frac{1}{4} \pi) \rbrace,
\end{equation}
the ingoing flux at infinity is given by 
\begin{equation}
F_{\infty} = \frac{ |\alpha |^2 }{2 \pi} = \frac{8}{ \pi^3 \omega^2}.
\end{equation}
The s-wave absorption probability is given by the ratio of the flux
across the horizon to the ingoing flux at infinity and hence
\begin{equation}
\sigma_{abs}^{S} = \frac{1}{4} \pi^2 \omega^4 R^4.
\end{equation}
Multiplying by $4 \pi /\omega^3$ to obtain the absorption
cross-section we find that
\begin{equation}
\sigma_{abs} = \pi^3 \omega R^4. \label{sem_clas}
\end{equation}
As we would expect, the absorption cross-section vanishes at zero
frequency. The absorption cross section for a minimally coupled scalar
in the associated five dimensional black hole under the assumption
that $r_{K} \ll r_{1}, r_5$ is \cite{MS}
\begin{equation}
\sigma_{abs} = \pi^3 \omega R^4 \frac{e^{\frac{\omega}{T_{H}}} -1}
{(e^{\frac{\omega}{2T_{L}}} - 1)(e^{\frac{\omega}{2T_{R}}} -1)},
\end{equation}
where $T_{L}$ and $T_{R}$ are the temperatures of the left and right
moving excitations respectively. 
In the limit that $r_K \rightarrow 0$, 
$T_{L}, T_{R} \rightarrow 0$, with the Hawking temperature defined as 
\begin{equation}
\frac{1}{T_{H}} = \frac{1}{2} \left( \frac{1}{T_{L}} + \frac{1}{T_{R}}
  \right ),
\end{equation}
we recover (\ref{sem_clas}) as required. As is by now well-known we
can also reproduce this result from the scattering cross sections of
BTZ black holes \cite{BTZ}: the cross section for a BTZ black hole is
\cite{HL_Y} 
\begin{equation}
\sigma = \pi^2 \omega R^2 \frac{e^{\frac{\omega}{T_{H}}} -1}
{(e^{\frac{\omega}{2T_{L}}} - 1)(e^{\frac{\omega}{2T_{R}}} -1)},
\end{equation}
where $R$ defines the asymptotic radius of curvature. One then takes
the same limit of the temperatures and multiplies by the volume of the
three sphere $2 \pi^3 R^3$ and divides by the the length of the circle
direction $2 \pi$ to obtain the string cross section. 
This limit corresponds to taking the zero mass, zero angular 
momentum BTZ black hole.  

In the near extremal limit
of the string, which we obtain by replacing $g_{tt} \rightarrow h
g_{tt}$ and $g_{rr} \rightarrow h^{-1} g_{rr}$ where
\begin{equation}
h = (1 - \frac{r_0^2}{r^2}),
\end{equation}
with $r_{0}$ the extremality parameter, 
the temperatures of the left and right moving excitations are given
by
\begin{equation}
T_{L} = T_{R} = T_{H} = \frac{r_{0}}{2 \pi R^2}.
\end{equation}
This indicates that the near extremal cross section is 
\begin{equation}
\sigma = \pi^3 \omega R^4 \coth(\frac{\omega}{4 T_{L}}), \label{near_ext}
\end{equation}
which is finite in the limit of zero frequency only for a black string far
from extremality.

\bigskip

From a supergravity point of view the dominant corrections to the absorption 
cross section arise from the matching about the self dual point
$r=R$. Following the approach of \cite{Kle_Has} we look for scalar
field solutions as power series in $\omega$: note that there does not
seem to be an exact solution to the wave equation as was found for
scattering within a 3-brane background in \cite{Gu_Ha}.  
In contrast to the D3-brane and M-brane calculations, 
we have not one but two dimensionless parameters
controlling the corrections, following from the presence of two scales
in the semi-classical geometry. This of course follows from the fact that the 
effective string preserves only one quarter of the supersymmetry. 
The two dimensionless parameters are
$\omega Q$ and $\omega R$, where $Q$ and $R$ are defined above.
In the limiting case $r_1 = r_5$, which has been distinguished as a special 
case several times in the literature, most notably in fixed scalar 
calculations \cite{CGKT}, \cite{KK2}, \cite{MMT3}, 
there is only one scale in the black string geometry. 

Since $Q$ is necessarily greater than $R$, 
in the region $r \ll R$ the right hand side of (\ref{can_eq1}) acts
as a small correction to the leading order solution. However, as we
approach the self-dual point $r=R$ the two
terms in the equation involving $\omega Q$ and $\omega R$ act as 
corrections to the leading order solution of the same order of magnitude.  
That is, in the near horizon
region $r < R$, the field equation for $f(y) = \phi(y)/y$ should be
written as 
\begin{equation}
\left [ y^2 f'' + y f' + (y^2 - 1) f \right ] = - \omega^2 Q^2 f -
\frac{(\omega R)^4}{y^2} f.
\end{equation}
For small $\omega Q$ and $\omega R$ the terms on the right hand side
act as small corrections even at the self-dual point. 
We look for a perturbative solution $f(y) = f_0(y) + f_1(y)$ where 
the leading order solution is 
\begin{equation}
f_0(y) = H_1^{(2)}(y),
\end{equation}
and $f_{1}$ satisfies the inhomogeneous equation 
\begin{equation}
\left [ y^2 f_{1}'' + y f_{1}' + (y^2 - 1) f_{1} \right ] = - \omega^2
Q^2 f_{0} - \frac{(\omega R)^4}{y^2} f_{0}.
\end{equation}
Since this is a second order equation we can simply write down the
solution for $f_1$ as
\begin{equation}
f_1(y) = \pi (\omega)^2 \int^{y} dx \lbrace \frac{Q^2}{2x} -
  \frac{\omega^2 R^4}{2 x^3} \rbrace f_0(x) \left [ J_1(x) Y_1(y) -
  J_1(y) Y_1(x) \right ]. \label{cor_eq}
\end{equation}
Of course $f_1(y)$ is ambiguous in the sense that one can add to it
any solution of the homogeneous equation. We can fix this ambiguity by
imposing the boundary conditions that all flux at the horizon is
infalling. When we follow the same procedure for $r > R$ we fix the
ambiguity by demanding that the solutions of the inner and outer
regions match to order $(\omega Q)^2 \ln (\omega R)$ in the transition
region. Analysis
of the matching in the transition region reveals that the dominant
correction to the flux ratio is of this order. There are also
corrections of order $(\omega Q)^2$ but these will be subleading for
small $\omega Q$.
With such a condition we find that the homogeneous part of the solution can be
taken to be $J_{1}(\omega r)/r$. Matching to higher order in fact 
requires that we also have a non-zero (but subleading) contribution to
the wave function from the homogeneous solution $Y_{1}(\omega r)/r$. 

Substituting the form for $f_0(y)$ and retaining only leading order terms 
in $(\omega Q)$, $\phi$ is then found to take the following form at small $y$
\begin{equation}
 \phi(y) = \frac{2 i}{\pi} \left ( 1 - \frac{1}{2}(\omega Q)^2 \ln(y) \right ).
\label{up_eq}
\end{equation}
There are also subleading correction terms of the form $(\omega Q)^2$ 
multiplied by powers of $y$; dominant corrections arise as we would expect from
the first term in (\ref{cor_eq}) only. If we rewrite this solution in 
terms of the variable $u = r/R$ we find that
\begin{equation}
\phi(u) = \frac{2 i}{\pi} \left ( 1 - \frac{1}{2}(\omega Q)^2 \ln(\omega R) + 
\frac{1}{2} (\omega Q)^2 \ln(u) \right ).
\end{equation}
We can repeat the same procedure in the region $r > R$ to find that
the leading order solution in the transition region is given by
\begin{equation}
\phi(\rho) = \phi((\omega R) u) = \omega \frac{\alpha}{2}
\left ( 1 + \frac{1}{2}(\omega Q)^2 \ln(\omega R) + \frac{1}{2} 
(\omega Q)^2 \ln(u) \right ). 
\end{equation}
We can compare these solutions at the self-dual point $u = 1$; since
$(\omega R)$ is much smaller than one, both $\rho$ and $y$ are small
in the matching region, and our perturbative expansions are valid. The
mismatch between these solutions requires that one take
\begin{equation}
\alpha = \frac{4 i}{\pi \omega} \left[1 - (\omega Q)^2 \ln (\omega
  R) \right ],
\end{equation}
which implies that the absorption cross section behaves as 
\begin{equation}
\sigma = \pi^3 \omega R^4  \left[1 + 2 (\omega Q)^2 \ln (\omega R)
\right ].
\end{equation}
It is straightforward to show that higher order corrections to the 
cross section are of the form one would expect  
\begin{eqnarray}
\sigma &=& \pi^3 \omega R^4 [ 1 + a_1 (\omega Q)^2 + a_2(\omega Q)^2
(\omega R)^2 + a_3 (\omega R)^4 + ..... \nonumber \\
&& \hspace{5mm} + \ln (\omega R) \left ( 2 (\omega Q)^2 + b_2 (\omega
  Q)^2 (\omega R)^2 + b_3 (\omega R)^4 + ... \right ) \\
&& \hspace{5mm} + (\ln (\omega R))^2 \left ( c_1  (\omega Q)^2 (\omega
  R)^2 + c_2 (\omega R)^4 + .... \right ) + ... ], \nonumber
\end{eqnarray}
where the ellipses indicate higher powers of $\omega Q$ and $\omega
R$. 

Note that this calculation implies that for 
scattering from the black string carrying momentum in the string
direction one should get corrections to the cross section of the same
type. That is, we expect the low energy cross section of the five
dimensional black hole to behave as
\begin{equation}
\sigma = A_{h} \left ( 1 + O((\omega Q)^2 \ln(\omega R)) \right ),
\end{equation}
where $A_h$ is the area of the horizon. Evidently for such an
expression to hold we need to assume that $r_{K} \ll Q$; in the region
$r_{K} \ll r \le R$ the metric will then be of the AdS form and this matching 
scheme will hold. If
$r_{K}$ is of the same order as $r_1$ and $r_5$ the matching of the scalar 
field wave function between regions is more subtle \cite{Do}. 

\section{The effective string model} \label{iv}
\noindent

We now consider the world sheet origins of the logarithmic corrections
to the cross section. Unlike the D3-brane case, we do not have a good
description of the action for the system at small $g$. The heuristic
model introduced in \cite{CM} and developed in \cite{DM1}, \cite{DM2},
\cite{CGKT} cannot produce the correct results for fixed scalars,
although this is not the case for minimal scalars. The refined model
discussed in \cite{HW} relies on the moduli space approximation, and 
the higher order couplings in which we are interested lie beyond the
scope of this approximation. With these problems in mind, we will use
the model of \cite{CGKT} and then investigate what input the
semi-classical results have on this model. 

So let us assume that the low energy excitations of the system are
described by the standard D-string action 
\begin{equation}
S_{D} = - T_{eff} \int d^2\sigma e^{-\Phi} \sqrt{ - \rm{det}(
  \gamma_{ij})} + ... ,
\end{equation}
with $\Phi$ the ten-dimensional dilaton and $\gamma$ the induced 
string frame metric on the world sheet defined by 
\begin{equation}
\gamma_{ij} = G_{MN} \partial_{i} X^{M} \partial_{j} X^{N},
\end{equation}
with $G$ the ten-dimensional string frame metric. As is usual we set
the world sheet gauge field to zero and we will choose the static 
gauge $X^{0} = x^{0}$, $X^{9} = x^{1}$. 
We are interested in the coupling of a minimally coupled scalar to the
string world sheet and will choose this scalar to be the six-dimensional
dilaton $\Phi_6$. Expanding out the action 
the relevant terms describing the coupling of the dilaton to the
world sheet are \cite{CGKT}
\begin{eqnarray}
S_{D} &=& - T_{eff} \int d^2x [ 1 + \frac{1}{2} \partial_{+}
  X^{m} \partial_{-} X_{m} - \frac{1}{2} \Phi_{6} (\partial_{+} X^{m} 
\partial_{-} X_{m})  \\
&& \hspace{10mm} + \frac{3}{16} \Phi_{6} (\partial_{+} X)^2 
(\partial_{-} X)^2 + \frac{1}{16} \Phi_6
  (\partial_{+}X^m)(\partial_{-}X_{m}) \left( (\partial_{+}X)^2 +
  (\partial_{-}X)^2 \right ) + ... ],  \nonumber
\end{eqnarray} 
where as usual $\partial_{+} = \partial_{0} + \partial_{1}$ and
$\partial_{-} = \partial_{0} - \partial_{1}$. In principle the index
$m$ runs over $1...8$ although we expect that only fluctuations
within the 5-brane are significant \cite{CGKT}, and hence we should
sum only over $m = 5..8$. There is a subtlety in the subleading terms:
to take account of two bosonised fermion fields $\varphi$ it seems that 
one should add these fields as follows \cite{CGKT}
\begin{equation}
(\partial_{+} X)^2 \rightarrow (\partial_{+} X)^2 + (\partial_{+}
\varphi)^2, 
\end{equation}
and similarly for the left derivatives. Such a correction is required
to obtain the correct normalisation for the fixed scalar cross section
in the case $r_1 = r_5$. 
Introducing canonically normalised scalar fields $\tilde{X}^m$ such that 
\begin{equation}
\tilde{X}^m = \sqrt{T_{eff}} X^m,
\end{equation}
and rotating to Euclidean signature $x^0 \rightarrow
i x^{0}$ we find the action becomes 
\begin{eqnarray}
S_{int} &=& - \int d^2\sigma [ T_{eff} - \frac{1}{2} (\partial
\tilde{X})^2 + \frac{1}{2} \Phi_{6} (\partial\tilde{X})^2 +
\frac{3}{16 T_{eff}} \Phi_{6} [(\partial\tilde{X})^2]^2 \nonumber \\\
&& \hspace{20mm} +  \frac{1}{8 T_{eff}} \Phi_{6} (\partial
\tilde{X})^2 \left( (\partial_{0}\tilde{X})^2 -
  (\partial_{1}\tilde{X})^2 \right) + ...], 
\label{int_ac}
\end{eqnarray}
where 
\begin{equation}
(\partial\tilde{X})^2 = \sum_{m} [( \partial_{0} \tilde{X}^{m})^2 + 
( \partial_{1} \tilde{X}^{m})^2 ].
\end{equation}
Hence we find that at linear order the dilaton couples to the world volume 
through an interaction of the form 
\begin{equation}
S_{int} = - \int d^2 x (\phi {\cal{O}}) = - \int d^2 x \phi 
\left [{\cal{O}}_2 + \frac{1}{T_{eff}} {\cal{O}}_{4} + ... \right ],
\end{equation}
where the subscripts to the operators indicate their conformal dimensions. 
The effective tension $T_{eff}$ has length dimension of minus two and  
as one expects one picks up 
factors of $1/T_{eff}$ as one increases the operator dimension. we are
interested in calculating the two point function of the operator
${\cal{O}}$: to leading order we can calculate using the infrared
limit. However, subleading corrections arise from the effect of
irrelevant perturbations which take the theory away from the
superconformal limit. 
Following the same type of analysis as in \cite{Kle_Has} one finds that the two
point function for the operator ${\cal{O}}$ is given by 
\begin{eqnarray}
\left < {\cal{O}}(x) {\cal{O}}(0) \right > &=& \int {\cal{D}} X
e^{-\int d^2 y \left [ {\cal{O}}_2 + \frac{1}{T_{eff}}
    {\cal{O}}_{4} \right ]} {\cal{O}}(x) {\cal{O}}(0);
\nonumber \\
&=& \int {\cal{D}} X e^{-\int d^2 y {\cal{O}}_2} {\cal{O}}(x)
{\cal{O}}(0) \left ( 1 - \frac{1}{T} \int d^2 z {\cal{O}}_4(z)
\right ); \nonumber \\ 
&=& \left < {\cal{O}}(x) {\cal{O}}(0) \left ( 1 -
    \frac{1}{T_{eff}} \int d^2 z {\cal{O}}_4(z) \right ) \right >;
\\
&=& \left [ \left < {\cal{O}}_2(x) {\cal{O}}_2(0)
  \right > - \frac{1}{T_{eff}} \int d^2 z \left < {\cal{O}}_2(x)
    {\cal{O}}_4(z) {\cal{O}}_2(0) \right > \right ]. \nonumber 
\end{eqnarray} 
The correlators on the right hand side are evaluated in the free gauge 
theory which is conformal, and all subsequent correlators are implicitly 
evaluated in the conformal theory. 
Note that we have implicitly gone to Euclidean signature which will be
used to simplify the correlator calculations. Terms like $\left <
  {\cal{O}}_2(x) {\cal{O}}_4(0) \right >$ vanish since only
operators of the same conformal dimension can have a non-vanishing two
point function.

\bigskip

To calculate the absorption cross section which follows from the leading order 
interaction term we could use the methods of \cite{DM1} taking the zero 
temperature limit. Since we also
wish to calculate the subleading corrections it is instructive to use
instead the methods used in \cite{Kl_Gub} and in \cite{Kle_Has}. At zero
temperature the analysis in fact becomes a great deal easier: 
absorption cross-sections corresponds up to a simple
overall factor to discontinuities of two point functions of certain
operators on the D-brane world-volume \cite{Kl_Gub}. 
Here we are considering here
minimally coupled massless particles normally incident on the string. 
If the coupling of the particles to the string is given by 
\begin{equation}
S_{int} = \int d^2 x \phi(x,0) {\cal{O}}(x),
\end{equation}
where $\phi(x,0)$ is a canonically normalised field evaluated on the
brane, and ${\cal{O}}$ is a local operator on the brane, then the
precise correspondence is 
\begin{equation}
\sigma = \frac{1}{2 i \omega} \hspace{2mm} 
{\rm{Disc}} \hspace{1mm} \Pi(k)|_{[k^0 = \omega; k = 0]},
\end{equation}
with $\omega$ the energy of the particle and 
\begin{equation}
\Pi(k) = \int d^2x e^{i k \cdot x} \left < {\cal{O}}(x)
  {\cal{O}}(0) \right >. 
\end{equation}
Disc $\Pi(k)$ is the difference of $\Pi(k)$ evaluated for $k^2 = \omega^2 +
i \epsilon$ and $k^2 = \omega^2 - i \epsilon$. The validity of this
expression depends on $\phi$ being a canonically normalised field. 

\bigskip
 
In Euclidean space the propagator for a scalar field $\tilde{X}$ is 
\begin{equation}
\left < \tilde{X}(x) \tilde{X}(0) \right > = \frac{1}{2 \pi} \ln(x),
\end{equation}
where $x^2 = (x_0^2 + x_1^2)$. Note that we are assuming that the string 
direction is infinite rather than compact. Then,
\begin{equation}
\left < \partial_{i} \tilde{X}(x) \partial_{I} \tilde{X}(0) \right >
= - \frac{1}{2\pi x^2} \left[ \delta_{iI} - \frac{2}{x^2} x_{i}
  x_{I} \right].
\end{equation}
From this we can deduce that the term giving the leading order
contribution to the absorption cross section is  
\begin{equation}
\Pi(x) = \left < :\frac{1}{2} (\partial\tilde{X})^2(x) : 
: \frac{1}{2} (\partial\tilde{X})^2(0): \right > = \frac{1}{8\pi^2 x^4},
\end{equation}
and Fourier transforming we find that 
\begin{equation}
\Pi(k) = \int d^2x \Pi(x) e^{i k \cdot x} = - \frac{k^2}{16 \pi}
\ln(k^2/\Lambda^2),
\end{equation}
where $\Lambda$ is an ultraviolet cutoff. 
The leading order absorption cross section is hence given by
\begin{equation}
\sigma = 4 \times \frac{\kappa_6^2}{2i \omega} \hspace{2mm}
\rm{Disc} \hspace{1mm} \Pi(s) = \frac{1}{4} \omega \kappa_{6}^2, 
\end{equation}
where the factor of four originates from the four scalars on the world
volume to which the scalar couples and the factor of $\kappa_6^2$ 
originates from the fact that the dilaton is not canonically
normalised in (\ref{6d_act}). 

Comparing the expressions for $R$ and
$\kappa_6^2$ in (\ref{eq_R}) and (\ref{eq_kap}) we see that there is a
discrepancy of $n_1 n_5$. However one expects that the
effective $\alpha'$ on the string world sheet is $\sqrt{n_1 n_5} \alpha'$
because of the fractionisation of the open string excitations \cite{Mald_5}. 
Hence the effective $\kappa_6^2$ on the world sheet
is indeed equal to $4 \pi^3 R^4$ and the string cross section agrees with the 
semi-classical calculation to leading order. 

\bigskip

The leading order correction to the cross section will be given by
\begin{eqnarray}
\delta\Pi(x) &=& - \int d^2 z \left < :\frac{1}{2} (\partial\tilde{X})^2(x):
  : \frac{1}{T_{eff}} {\cal{O}}_4(z): 
:\frac{1}{2} (\partial\tilde{X})^2(0): \right >; \nonumber \\
&=& \frac{3}{128 \pi^4 T_{eff}} \int d^2z \frac{1}{z^4 (x-z)^4},
\end{eqnarray}
where the form of ${\cal{O}}_4(z)$ follows from (\ref{int_ac}).
In momentum space, 
\begin{equation}
\delta\Pi(k) = \int d^2 x \Pi_1(x) e^{i k \cdot x} = \frac{3}
{512 \pi^2 T_{eff}} k^4 (\ln(k^2/\Lambda^2))^2.
\end{equation}
Then the correction to the absorption cross section is given by
\begin{equation}
\delta \sigma = 16 \times \frac{\kappa_6^2}{2 i \omega} \hspace{2mm}
{\rm Disc} \hspace{1mm} \delta \Pi(k),
\end{equation}
where the factor of $16$ arises from the fact that four
scalars contribute. We expect
there to be two bosonised fermions contributing to the subleading
interaction term in the action (\ref{int_ac}) as well as the four bosons: 
when we calculate cross sections for the fixed scalars we need to
include them to obtain agreement for the cross section in the case
$r_1= r_5$ \cite{CGKT}. However, the terms arising from the fermions 
do not contribute to the subleading term in the dilaton cross section. 
We find that 
\begin{equation}
\delta\sigma = \pi^3 \omega R^4 \left ( \frac{3 \omega^2 }{2 \pi T_{eff}} \ln
   (\omega/\Lambda) \right ).
\end{equation}
To compare this with the supergravity calculation we should take the
cutoff to be at $\Lambda = 1/R$. As in the D3-brane
calculations, the natural cutoff on the world sheet is
$1/\sqrt{\alpha'}$ but the difference between the cutoffs gives a
contribution to the cross section of the form 
\begin{equation}
\delta \sigma \propto \frac{\omega^3 R^4}{T_{eff}} \ln (R/\sqrt{\alpha'}).
\end{equation}
Hence the difference between these cutoffs
contributes only to the first non-logarithmic correction  
term in the cross section which we have not calculated. 

One then has to decide what value one should use for the effective
string tension. Analysis of the entropy and temperature of
near-extremal five branes leads to an effective string of tension $1/2
\pi r_5^2$ \cite{Mald_5}. Extending these methods to the case $r_1
\sim r_5$ implies that \cite{Gubs_2}
\begin{equation}
T_{eff} = \frac{1}{2\pi Q^2}. 
\end{equation}
However, most of the scattering calculations do not depend on the
effective tension or require $r_1 = r_5$, and so there is an ambiguity
in the value one should take for the effective tension. In fact it was
shown in \cite{Gubs_2}, \cite{Mat_hur} that one should choose the value   
\begin{equation}
t_{eff} = \frac{1}{2 \pi R^2}
\end{equation}
to obtain the correct scaling properties of cross sections of higher partial 
waves of minimal scalars from the effective string model. This is also
the value found in the analysis of \cite{HW}. This
ambiguity illustrates a problem of the effective string model: in this
zero temperature limit, there
are two scales in the geometry but only one of these scales appears in
the effective string action. For the general near extremal five
dimensional black hole, there are four scales in the semi-classical
geometry $r_{0}$, $r_{K}$, $Q$ and $R$, where $r_{0}$ is the
extremality parameter and $r_{K}$ is related to the Kaluza-Klein charge. 
In the effective string model, however, there are only three length
scales, given by $T_{eff}$, $T_{L}$ and $T_{R}$. 

\bigskip

It is interesting to note that if one
chooses the first value for the tension then our string correction
is a factor of $3/2$ greater than the semi-classical result. One
should not be worried by such a numerical discrepancy, since 
the semiclassical solution is valid in the 
regime $g n_{1}, g n_5 \gg 1$, whereas the perturbative effective 
string calculation is valid in the region $g n_1, g n_5 \ll 1$. 
We can only reliably compare the cross-sections in the
decoupling limit for which we cannot calculate perturbatively on the
string world sheet.

Even though the first correction agrees in form, if not coefficient, 
with that calculated semi-classically the effective string model cannot
then reproduce the next order correction of the form 
\begin{equation} 
\omega R^4 \left[ \left( \ln(\omega R) \right )^2 (\omega Q)^2 (\omega
  R)^2 \right ],
\end{equation}
since the scale set by $R$ does not appear in the effective string
model, except as an ultraviolet cutoff in the logarithmic terms. 

For finite temperature, we will have to use the finite temperature
Green's functions to calculate the correction to the cross
section. One can follow an approach similar to that given in
\cite{Gubs_1} to calculate this correction. The leading order term
follows straightforwardly from \cite{Gubs_1} and reproduces the 
$r_{K} \rightarrow 0$ limit of the result of \cite{MS}. One can also
show that the functional dependence of the leading order correction 
behaves as
\begin{equation}
\frac{\omega^3 R^4}{T_{eff}} \coth \left( \frac{\omega}{4 T_{L}}
\right ),
\end{equation}
for $r_{0} \ll R$ as one would expect from (\ref{near_ext}).  

\bigskip

The appearance of two scales in the semi-classical geometry but only
one scale in the effective string model is also related to the
discrepancy in calculations of fixed scalar cross sections for $r_1
\ne r_5$. In the limit of $T_{L} = T_{R}$ the discrepancy in
functional dependence of the leading order cross sections disappears:
that is, one obtains the same functional form for the cross section 
from $(1,3)$, $(2,2)$ and $(3,1)$ operators. Given the conformal
dimension of these operators, it is easy to see that the leading order cross
section for both fixed scalars behaves as 
\begin{equation}
\sigma \sim \frac{1}{T_{eff}^2} \omega^5 R^4.
\end{equation}
However, the precise results calculated semi-classically are
\cite{KK2}, \cite{MMT3} 
\begin{equation}
\sigma = \frac{9 \pi^3 \omega^5 R^{12}}{64 ( Q^2 \pm \sqrt{Q^4 -
    3R^4})^2},
\end{equation}
where the sign depends on which of the two fixed scalars we are considering.

\section{The CFT correspondence} \label{v}

Having considered the world sheet interpretation of subleading effects
in the $g \rightarrow 0$ limit, it is interesting to consider the
calculation of the absorption cross-section from
the AdS-CFT correspondence in the limit of large $g n_1$, $g n_5$. 
In the region $r \ll R$ the geometry of
the black string is that of (\ref{str_met}) and we will assume in all that 
follows that the string direction is not compact. 

The AdS-CFT correspondence implies
that a massless minimally coupled scalar couples at leading order 
to an operator of conformal dimension two. The two point function 
of the operator ${\cal{O}}$ behaves as 
\begin{equation}
\left < {\cal{O}}(k) {\cal{O}}(q) \right >  \hspace{1mm} \propto 
\hspace{1mm} \frac{R^4}{\kappa_6^2} k^2 \ln (kR)^2 \delta^2(k+q),
\end{equation}
where $\kappa_6$ is the six-dimensional gravitational
constant. One could obtain this form directly from Fourier transforming 
the two point functions of \cite{W_1} and \cite{FMMR}, but to fix the 
normalisation it is convenient to follow an 
analysis similar to that of \cite{PoKl}
{\footnote{Since the field is massless we do not need to
  worry about the correction factor discussed in \cite{FMMR}.}}. 
Starting with an action for the scalar field of the form 
\begin{equation}
S = \frac{1}{2 \kappa_{6}^{2}} \int d^6x \sqrt{-g} \left ( \frac{1}{2} 
(\partial \phi)^2 \right ), \label{act_pcn}
\end{equation}
and substituting for the metric (\ref{str_met}) we find that 
\begin{equation}
S = \frac{\pi^2 R^4}{ 2 \kappa_6^2} \int d^2 x \left [ \phi \frac{1}{z}
\partial_z \phi \right ]^{\infty}_{R},
\end{equation}
where we have introduced a cutoff at radius $r=R$ and $x = (t,x)$. 
The equation of motion for the field $\phi$ is 
\begin{equation}
\left [ z \partial_z \frac{1}{z} \partial_{z} + \eta^{ij}
  \partial_{i} \partial_{j} \right ] \phi = 0,
\end{equation}
where $i,j = 1,2$. Finiteness of the
action requires that $\phi$ must vanish in the limit that 
$z \rightarrow \infty$ and the appropriate form for $\phi$ is then
\begin{equation}
\phi(x,z) = \frac{1}{(2\pi)^2}
\int d^2k \lambda_k e^{i k \cdot x} \left (\frac{ z K_{1}(kz)}{R
  K_{1} (kR)} \right),
\end{equation}
where $K_1$ is the modified first Bessel function. 
Note that we have normalised the scalar field so that it takes a value
of one on the boundary $r=R$. Substituting into the action we find to
leading order 
\begin{equation}
S = - \frac{\pi^2 R^4}{4 \kappa_6^2} \int d^2 k d^2 q  \lambda_k
\lambda_q \left ( \frac{1}{(2 \pi)^2} \delta^2(k+q) \right ) k^2 \ln (kR)^2,
\end{equation}
from which it is apparent that the two point function is 
\begin{equation} 
\left < {\cal{O}}(k) {\cal{O}}(q) \right >  = -
\frac{\pi^2 R^4}{2 \kappa_6^2} k^2 \ln (kR)^2 \left ( \frac{1}{(2
    \pi)^2} \delta^2(k+q) \right ).
\end{equation}
Defining $s = - k^2$ and letting 
\begin{equation}
\Pi(s) = \frac{\pi^2 R^4}{2 \kappa_6^2} [s \ln(- R^2 s)],
\end{equation}
then the absorption cross section can be inferred from the
discontinuity as one crosses the positive real axis in the $s$-plane
of $\Pi(s)$. That is, the cross section behaves as
\begin{equation}
\sigma = \frac{\kappa_6^2}{i \omega} {\rm{Disc}} (\Pi(s))|_{[k^0 =
  \omega, k = 0]},
\end{equation}
where the factor of $\kappa_6^2$ arises from the fact that $\phi$ is
not canonically normalised in (\ref{act_pcn}). Hence we find that 
\begin{equation}
\sigma = \pi^3 \omega R^4, 
\end{equation}
which is the same as the semi-classical result (\ref{sem_clas}) and
the leading order string result as expected. 

\bigskip

To look for subleading corrections to this cross section we need to
postulate what the effective action for the conformal field theory 
is in the large $g n_1$, $g
n_5$ limit. Looking at the form of the semi-classical 
corrections one sees that to reproduce this result one must 
have a correction to the two point function coming from a term of the form 
\begin{equation}
\int d^2 z \left < {\cal{O}}_2(x) {\cal{O}}_4(z) {\cal{O}}_2(0) \right > \sim 
\frac{1}{T} \int d^2 z \frac{1}{z^4 (x-z)^4},
\end{equation}
where $T$ has length dimension of minus two. Then one infers that the
correction to the cross section is 
\begin{equation}
\delta \sigma \propto \kappa_6^2 \omega \left ( \omega^2
T^{-2} \ln (\omega/\Lambda) \right ).
\end{equation}
Since the only length scale in the CFT is $R$ we should take 
$T \propto 1/R^2$ which gives the correction to the cross-section as 
\begin{equation}
\delta \sigma \propto \omega^3 R^6 \ln (\omega R),
\end{equation}
where we have also taken the cutoff at $1/R$. 
The conformal field theory correction does not then agree in normalisation with
that of the semi-classical calculation. Such a discrepancy is unsurprising
given that the length scale $Q$ does not appear in the boundary
conformal theory. 

The fixed scalars couple to operators of dimension four, and one can
hence find that the absorption cross section for both of them is
given by
\begin{equation}
\sigma = \frac{1}{64} \pi^3 \omega^5 R^8,
\end{equation}
which again does not agree with the semi-classical calculation. The fixed
scalars are scattered in the asymptotically flat part of the geometry 
which is not described by the boundary theory. It is interesting to
note that in the CFT approach there is an ambiguity in the operators to
which the fixed scalars couple which affects the finite temperature
result. We know that the operator has
dimension four, but cannot fix whether it is $(3,1)$, $(2,2)$ or
$(1,3)$ without further analysis. It is of course natural to assume that the
operator is of the $(2,2)$ type, as was done in \cite{Teo}, 
since this reproduces the functional form of the semi-classical results. 
The same ambiguity will appear in finite temperature calculations of
scattering from D3-branes and M-branes. To fix the ambiguities one
needs to know the origins of the terms in the boundary
conformal field theories. 

\section{Conclusions} \label{vi}
\noindent

Our results demonstrate the limited applicability of the conjectured
AdS-CFT correspondence in the calculation of scattering in 
asymptotically flat systems using CFT methods. Although we can
calculate the leading order absorption rates using the properties of
the asymptotic near horizon geometry, subleading corrections depend
not just on the boundary theory but also on the bulk theory. That is,
the asymptotically flat part of the geometry will determine the leading
order corrections in the semi-classical absorption rate. The boundary
CFT does not have information about the full geometry and hence cannot
give the correct answer for scattering in the asymptotically flat
geometry. 

If one has a non-minimally coupled particle in the semi-classical
geometry, then in many important cases the functional dependence of
the scattering rate seems to be determined solely by the behaviour in 
the near horizon region. Scattering in the asymptotically flat region
simply corrects the rate by factors depending on the parameters of the
near horizon geometry. This will mean that one can predict the
functional dependence of the scattering rate from the coupling to the
near horizon geometry, and hence from knowing the conformal dimension
of the operator in the CFT to which the particle couples. The
normalisation of the cross section could not however be predicted. 

Since the Maldacena conjecture relates type IIB theory in the anti-de Sitter
background to a dual conformal field theory, there is of
course no
reason why this conformal field theory should reproduce results in an
asymptotically flat spacetime in which only the near horizon geometry
is of the anti-de Sitter form.  
It would hence be interesting to investigate subleading effects in the
scattering of massive and massless scalars in the six-dimensional
geometry consisting of the BTZ black hole times a three sphere: one
would expect that these could be reproduced by the dual 
conformal field theory. The supergravity analysis of subleading
corrections is in this case more subtle because of the timelike
boundary at infinity. 

\bigskip

Our analysis also illustrates various problems in trying to reproduce the
semi-classical results using an effective string model. 
If the effective string model is to reproduce sub-leading
effects in the semi-classical geometry, one needs to incorporate
into the model the four length scales that generally determine the black hole
geometry. 
Of course there is no reason why the subleading effects calculated at 
$g n_1, g n_5 \gg 1$ should be reproduced by an effective string model valid
only for small $g$.

\end{document}